# LEARNING ABOUT COMETS FROM THE STUDY OF MASS DISTRIBUTIONS AND FLUXES OF METEOROID STREAMS

**Josep M. Trigo-Rodríguez[1,2], and Jürgen Blum[3]**

[1] Institute of Space Sciences (CSIC), Carrer de Can Magrans, s/n, Campus UAB,
08193 Cerdanyola del Vallés (Barcelona), Catalonia, Spain. E-mail: trigo@ice.csic.es
[2] Institut d'Estudis Espacials de Catalunya (IEEC), Edif.. Nexus,
c/Gran Capità, 2-4, 08034 Barcelona, Catalonia, Spain
[3] Institut für Geophysik und extraterrestrische Physik, Technische Universität Braunschweig,
Mendelssohnstr. 3, 38106 Braunschweig, Germany.

**Abstract**:

Meteor physics can provide new clues about the size, structure, and density of cometary disintegration products, establishing a bridge between different research fields. From meteor magnitude data we have estimated the mass distribution of meteoroids from different cometary streams by using the relation between the luminosity and the mass obtained by Verniani (1973). These mass distributions are in the range observed for dust particles released from comets 1P/Halley and 81P/Wild 2 as measured from spacecraft. From the derived mass distributions, we have integrated the incoming mass for the most significant meteor showers. By comparing the mass of the collected Interplanetary Dust Particles (IDPs) with that derived for cometary meteoroids a gap of several orders of magnitude is encountered. The largest examples of fluffy particles are clusters of IDPs no larger than 100 μm in size (or $5\times10^{-7}$ g in mass) while the largest cometary meteoroids are centimeter-sized objects. Such gaps can be explained by the fragmentation in the interstellar medium or in the atmosphere of the original cometary particles. As an application of the mass distribution computations we describe the significance of the disruption of fragile comets in close approaches to Earth as a more efficient (and probably more frequent) way to deliver volatiles than direct impacts. We finally apply our model to quantify the flux of meteoroids from different meteoroid streams, and to describe the main physical processes contributing to the progressive decay of cometary meteoroids in the interplanetary medium.

1. INTRODUCTION.

Comets are fragile objects formed in remote regions of the protoplanetary disk from the accretion of primordial materials made of dust grains, organics and ices (Oró, 1961; Brownlee et al., 2006; Nittler et al., 2019). When a comet approaches the Sun, its surface is heated and volatile species sublimate, essentially as described by Whipple (1950, 1951). A direct consequence is the continuous release of particles (Gundlach et al. 2020) that are accelerating away from the nucleus through drag from the outflowing gas (Skorov et al. 2018). It is well known that the meteoroid ejection process is mass dependent (see e.g. Jenniskens, 1998). Our knowledge on the release of dust from comets has evolved significantly thanks to *Stardust* (NASA) studies of 81P/Wild 2 (Brownlee et al., 2006) and also to the comprehensive studies made by ESA's *Rosetta* spacecraft of the comet 67P/Churyumov-Gerasimenko (Rotundi et al., 2015). The coma surrounding a comet is formed by evaporated gas and outflowing dust, and we know that clouds of larger grains orbit the nucleus from previous perihelia. The solar radiation pressure affects the movement of the smallest grains, while the largest ones remain closer to the nucleus as a consequence of the influence of the parent body's gravity. The first "dirty ice ball" comet model was extremely naïve in comparison. Recent studies of comets have revealed a far more complex comet structure, reminiscence of being objects that accreted in distant regions, but experienced encounters with other objects in a wide range of sizes. Many comets exhibit bi-lobed structures as a consequence of low relative velocity collisions that piled several blocks together. Comets became heterogeneous and complex since their accretion, born within remote protoplanetary disk regions rich in organics and ices, while others were probably depleted of them (Brownlee et al., 2006). As a consequence of the inherited heterogeneity and the variable *dust:organics:ices* ratios, the particular ejection processes of meteoroids from comets and their mass distributions observed when crossing as dust trails or evolved meteoroid streams are quite different from that produced during asteroidal impacts (Williams & Murad, 2002). All these meteoroids are subjected to different physical processes that make them to fragment to produce dust (Grün et al., 2019). Many observational techniques are currently available for the scientific study of meteoroids and dust, and we try here to compare their results.

In this paper, we will focus on the study of the mass distributions, bulk physical properties and strength of cometary meteoroids in order to gain insight into the accretionary processes, and make an exercise of quantifying the magnitude of the flux over time and the delivery of water and moderately volatile species, like e.g. Na. Greenberg & Li (1999) and Greenberg (2000) remarked the important difference between cometary particles and Interplanetary Dust Particles (IDPs) and hypothesized about their highly-porous structures, an aspect confirmed by the *Stardust* and *Rosetta* missions (Bronwlee et al., 2006; Heinisch et al., 2015; Levasseur-Regourd et al. 2018). It is obvious that most collected IDPs have suffered moderate heating, in space or during their atmospheric entry, so could have depleted of volatile interstitial materials (see e.g. Flynn, 2005, 2020). We try to gain insight into the physical properties of cometary matter by comparing the results obtained from meteor observations with the data from other techniques and space missions. An example is the data obtained from IDPs collected in the Earth's lower stratosphere (see e.g. Rietmeijer, 1998, 2002). Meteor science can have an added value when discussing the physical processes that these fluffy particles undergo in the interplanetary medium and when penetrating the terrestrial atmosphere (Vida et al., 2021). Cometary particles are assumed to constitute a significant component of the IDP flux onto the Earth, although the precise fraction is unknown, and the identification difficult (Brownlee, 1985; Jeffers et al., 2001; Rietmeijer, 2002; Wooden, 2002; Koschny et al., 2019).

2. METHODOLOGY.
2.1. MASS DISTRIBUTIONS OF METEOROID STREAMS

Compilations of the mass influx reaching the Earth from annual showers and meteor storms are scarce, except for some major streams (Hughes, 1974a,b; Jenniskens, 1994, 1995; Ceplecha et al., 1998; Brown, 1999; Trigo-Rodriguez et al., 2001, 2004a). By using all available observational data, we analyzed the meteoroid mass distribution for different cometary streams. We used the dependence between the maximum meteor magnitude and the meteoroid mass obtained by Verniani (1973). Trigo-Rodríguez et al. (2001, 2004b, 2013) already presented a procedure to compute the incoming mass from the observed Zenithal Hourly Rate (ZHR) and magnitude distributions. Here we apply this method in order to create mass distributions by comparison with the dust component ejected from different comets. To obtain reliable magnitude distributions, we checked the modeled data with that derived from the Visual Meteor Database of the International Meteor Organization IMO (IMO, 2000) and the Visual and Photographic Database of the Spanish Photographic Meteor Network (SPMN). In fact, visual observations have played an important role in the characterization of the rate profiles for the major showers on a year-to-year basis (Jenniskens, 1994, 1995). Carefully corrected visual analyses obtained by using the same procedures in the framework of the IMO show the interest of global observational amateur programs (see e.g. Rendtel & Arlt, 1997; Arlt, 1998, 1999). However, estimations of the magnitude distributions by visual observers can introduce important biases. Fortunately, during the last decades photographic, CCD and video observations of meteors have been able to provide detailed information on meteoroid mass distributions, just where the visual technique can be more subjective (Brown et al., 2002; Trigo-Rodriguez et al., 2004a).

Our approach consists of deducing the flux number density of meteor streams by using the ZHRs recorded in historical visual observations or in photographs (Jenniskens, 1995; Brown, 1999; Trigo-Rodríguez et al., 2001). Then, our theoretical approach is built on the basis of observational data available in the literature. Consequently, the mass distributions derived here are consistent with the observed ZHR and mass population for every stream. We can also estimate the increase in the mass influx during particular high-activity levels of meteor showers (like e.g. meteor storms).

The ZHR is the number of shower meteors that an observer would see in one hour under clear skies, the radiant at the zenith and a faintest visible star in the field of view equal to +6.5. Usually visual ZHR is obtained by using the following equation:

$$ZHR = \frac{C \cdot N \cdot r^{6.5-Lm}}{T \cdot (\sin \theta)^{1,4}} \qquad (1)$$

where $C$ is a correction for the perception of the observer relative to an average observer, $N$ is the number of shower meteors recorded in $T$ hours of effective time, $Lm$ is the limiting stellar magnitude and $\theta$ is the elevation of the shower radiant. In order to obtain the ZHR, we need to know the number of particles in each magnitude class. Such information is implicit in Eq. (1) in the term containing the population index $r$. The population index is the ratio between the number of meteors in magnitude class $m$ to class $m-1$ for all magnitude ranges in the observational interval.

$$r = \frac{N(m)}{N(m-1)} \qquad (2)$$

We used the population index in order to reconstruct the magnitude distributions of the meteor showers. Then, the observed ZHRs are estimated by taking into account the probability of perception for each magnitude class for a standard sky limiting magnitude of +6.5,

$$N_m = r^{m-M_o} \qquad (3)$$

The magnitude origin of the distribution ($M_o$) is considered a free parameter to obtain the ZHR values. However, we took into account that the modeled distributions fit the observed ones. In order to simulate reliable ZHRs, we searched in the literature for proper $r$ and $M_o$ values that are the basis of our model (IMO, 2000; Jenniskens, 1994, 1995; Trigo-Rodríguez et al., 2001). In the literature, an increase in the meteor activity observed from a meteoroid stream can be called outburst or storm. Usually a meteor outburst is considered a detectable increase in the level of activity from a meteor shower that produces ZHR<1000, while if this level is exceeded, it is called a meteor storm. This terminology will be followed hereafter. During outbursts and storms, the population index $r$ may deviate from that of the average annual stream as was noted previously (see e.g. Jenniskens, 1995). Typically, these meteor activity increases are produced by Earth's encounter with a cometary dust trail characterised for having meteoroids with lower release ages than the members of the annual stream. In fact, this population index can change along the whole magnitude range, although over the visual magnitude range, it can be assumed constant to a first approximation (Ceplecha, 1992). In our simple calculations, we usually assumed $r$ to be constant during the full activity period, but when the evolution of the population index during meteor storms was known, we incorporated it in order to improve the integration. The ZHR in each case was obtained by correcting the observed magnitude distribution for by a standard probability function $P(m)$ (Koschack & Rendtel, 1990)

$$N(m) = T(m) \cdot P(m). \qquad (4)$$

Here, *N(m)* and *T(m)* are the observed and real number of meteors for each magnitude class *m*. We used for the average perception function P(Δm) a fitting function formed by the hyperbolic tangent (Wu, 2005):

$$P(\Delta m) = 0.5 + 0.505 \tanh(0.66\Delta m - 0.013\Delta m^2 - 2.43) \qquad (5)$$

Figure 1 shows a typical mass distribution characteristic of a Leonid storm compared with the number of meteors that would be observed visually. It is clear that the visual perception for different magnitude classes depending on the particular limiting magnitude conditions must be taken into account for studies of meteoroid influx. In order to compute the global incoming mass, we operate as follows. We take the population index value of the particular meteor shower under study from the literature. Then, we proceed to create a mass distribution by assuming different $M_o$. When the modeled mass distribution produces the ZHR level given in the literature, the solution is achieved. Finally, the program integrates the mass from the number of meteoroids participating in every magnitude class. Basically, we multiply the number of meteors of each magnitude class by the estimated mass for every class by using the equation (Verniani, 1973):

$$0.92 \cdot \log m = 24.214 - 3.91 \cdot \log V_g - 0.4 \cdot M_v \qquad (6)$$

where *m* is the meteoroid mass in grams, $V_g$ is the geocentric velocity given in cm/s and $M_v$ is the magnitude of the produced meteor. Using this approach, we obtained the mass contributing to all magnitude classes from the meteors observed from a given location.

The final mass entering into the terrestrial atmosphere is finally calculated considering that the ZHR is given by definition for an effective area of the atmosphere covered by one visual observer. This effective area is ~128 times smaller than that covered by the terrestrial atmosphere globally (Trigo-Rodríguez et al., 2001; 2013). Consequently, we multiply this factor with the previously obtained mass in order to estimate the global mass entering into Earth's atmosphere. Considering the uncertainties associated with the population index and $M_o$, we estimated the error involved for well-studied storms to be on the order of a factor 1σ. As an application of the model, we quantify the products of the disruption of fragile carbonaceous asteroids and comets in close approaches to Earth as a more efficient (and probably more frequent) way to deliver volatiles than direct impacts.

We also note that a previous study can be used to quantify the validity of the numerical integration. We can use the mass-frequency distribution of micrometeoroids studied by Love and Brownlee (1993). They computed the mass per mass decade hitting the upper atmosphere from the impact craters left by this population of micrometeoroids hitting the *Long Duration Exposure Facility*. Interestingly, Love and Brownlee (1993) obtained a mass-frequency distribution exhibiting a peak in the impacting mass per mass decade occurring at ~$10^{-5}$ g. Then, to produce this distribution there must be a change in the slope of the mass distribution of small meteoroids at that specific mass (see Figure 3 of Love and Brownlee, 1993).

Consequently, if we consider the "computed distribution" shown in Figure 1 of this paper, the mass influx to Earth from each stream could be overestimated at some extent. In any case, we note that such constant population index distribution is usually applied in meteor observations to correct the faintest meteors for an eyewitness observer that misses a part of the faintest meteors when the shower is not observed under good sky conditions (limiting magnitude +6.5). An interesting output is the fact that, when we apply such type of correction, the final distribution of masses overestimates the mass arrived from the smallest particles. For simplicity, we decided to keep a constant population index correction factor that is widely accepted to infer meteoroid fluxes in meteor studies. In any case, we think that a future correction could be considered to better fit Love and Brownlee (1993) findings. The best way for doing so could be to increase the observational data of meteor streams flux for faint meteors using optical instruments (e.g. binoculars, or CCD cameras) to infer possible changes in the population index for faint meteors.

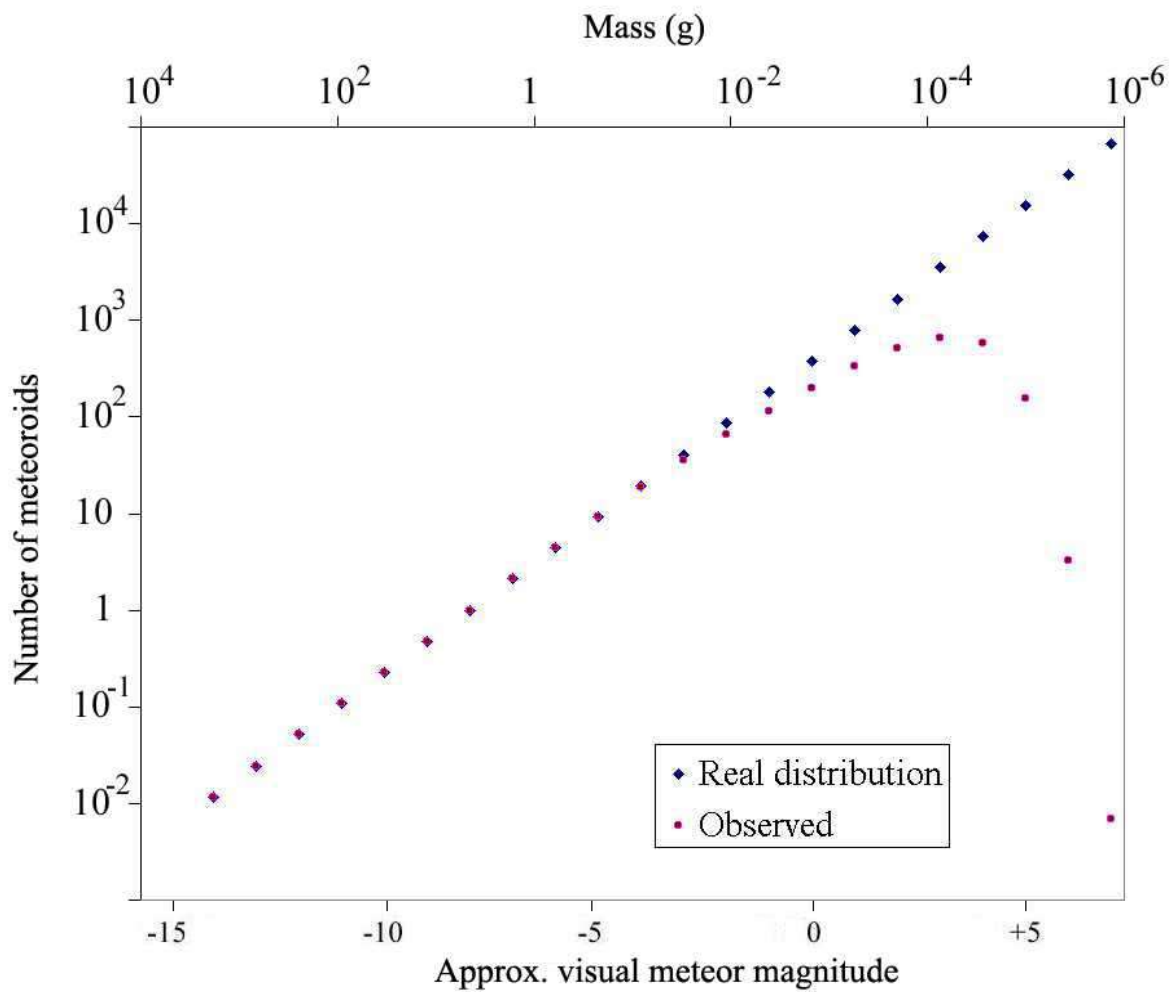

Figure 1. Characteristic mass distribution of a Leonid storm compared with one observed visually under perfect conditions. A constant population index of r=2.1 is assumed for all magnitude classes. (Trigo-Rodríguez et al., 2001)

## 3. RESULTS.

### 3.1. MASS DISTRIBUTIONS OF COMETARY STREAMS.

From the simulated mass distributions, we are able to estimate the global mass reaching the Earth from some of the most important annual meteor showers and storms (Table 1). Usually cometary streams with high relative velocities are famous because of their vusual appearances, e.g. the Perseids or Leonids, but are they delivering significant amounts of cometary material? In order to answer this question, we applied our model to estimate the incoming flux for both streams. For the Perseid meteoroid stream, our model shows an annual mass delivery of about 150±50 kg, while the mass delivery of the Leonid stream is two orders of magnitude lower. These values are incremented significantly during those years in which a meteor outburst or storm occurs. In a few hours, both streams are able to deliver tens of kilograms of cometary material to the upper atmosphere. We note that the extraordinary 1998 Leonid outburst was rich in large meteoroids contributing significantly to the incoming mass. During an outburst of about 20 hours, the Earth received ~2 metric tons of material released by comet 55P/Tempel-Tuttle. This is a good example of how the population index can affect the mass distribution and, consequently, the incoming flux. Usually, meteoroid streams are composed of extremely small particles that contribute only little to the delivery of cometary material. However, during short-time encounters of the Earth with young and dense dust trails, the influx can be much higher than that associated with the annual stream. The main reason is that young cometary dust trails are to a large degree composed of centimeter-sized meteoroids. This may indicate that long-time exposures to the interplanetary medium affect the population of large meteoroids.

Cometary streams with low relative velocities, like the October Draconids (Giacobinids) and the Pons-Winneckids, are efficiently delivering substantial amounts of cometary matter to the Earth. Another efficient contributor is the Taurid meteoroid complex that is associated with comet 2P/Encke. It is deduced that streams with low relative velocity encounters with the Earth efficiently deliver cometary matter to the terrestrial atmosphere as is the case for comets 26P/Grigg-Skjellerup and 73P/Schwassmann-Wachmann 3 (Messenger, 2002). Our calculations provide a first approach to the amount of cometary material deposited into the terrestrial atmosphere (Table 1). Depending on the different geometrical conditions and the number densities detected in these streams, the deposited mass can change by several orders of magnitude between low-velocity and high-velocity cometary streams.

In order to derive the particle masses, we used the photometric magnitude of the respective meteors produced when entering the terrestrial atmosphere (Verniani, 1973). In Figure 2a we plotted the observed mass distribution of particles in the coma of comets 1P/Halley and 81P/Wild 2 measured by instruments on board the *Giotto* and *Stardust* spacecrafts (Mc Donell et al., 1987; Green et al., 2004; Hörz et al., 2006). For comparison, we show the mass distributions of particles associated to three different comets whose dust trails are intercepted by the Earth: 1P/Halley (producing the Orionids), 55P/Tempel-Tuttle (Leonids) and 109P/Swift-Tuttle (Perseids). For the Leonid meteoroids, we show two distributions with different slopes. The one called "Leonids" shows the typical slope and dust distribution observed for this meteoroid stream. However, the "1998 Leonid outburst" possessed a different slope. The corresponding cometary activity, rich in large particles, ejected the dust trail particles in the year 1333 under a 5/14 mean motion resonant orbit with Jupiter (Asher et al. 1999). The Leonid dust distribution exhibits a larger slope than the one observed in the 1998 Leonid outburst. This feature can be interpreted as an aging effect, e.g. the number of small particles increases with age and consequently the slope also increases. This effect is widely observed in meteoroid streams that exhibit the brightest meteors during the peaks of meteor activity (see e.g. Jenniskens, 1994; 1998). It is interesting to remark that the mass distribution of large meteoroids of the 1998 Leonid outburst fits well to the dust distribution observed for young particles inside the coma of comet 81P/Wild 2. Another observation is that the Perseid mass distribution is similar to the one determined for the Leonids. Finally, the Orionid mass distribution fits well the one observed in the coma of 1P/Halley (as we expect for the meteoroid stream produced by the same parent body). However, a different cut with the x-axis and slope would indicate that the particles constituting the meteoroid stream are formed by smaller samples than those forming the coma, such as we expect from progressive fragmentation due to aging. These features show the consistency of the modeled meteoroid mass distributions derived from the approach of Verniani (1973).

In Figure 2b we compare the mass distributions of dust released from comets Arend Roland, Bennet and Seki-Lines as derived from remote observations (data from Delsemme, 1983) with those of dust measured in the comae of comets 1P/Halley and 81P/Wild 2. We see that the mass distribution of particles from 1P/Halley is very similar to the one observed for c/Seki-Lines. A similar concordance is observed for 81P/Wild 2 and c/Arend-Roland. However, the 81P/Wild 2 mass distribution is richer in large particles and we also notice that the mass distribution is far from a power law. In fact, it was found using the forward-scatter technique that the observed meteor streams appear to have very unstable populations (Cevolani and Gabucci, 1996).

Figure 2b also shows an interesting feature that requires additional discussion. Remote observations of cometary comae exhibit mass distributions with a clear cut-off for very small particles (with masses $\sim 10^{-12}$ to $10^{-14}$ g). We also made a literature search for evidence of the presence of some cut-off in the magnitude distributions of meteoroid streams intercepted by the Earth. We looked for data obtained from four different observing techniques: visual, photographic, video and radar. The existence of a cut-off for large particles is well known because cometary ejecta must escape the gravitational field of the cometary nucleus. As the ejection is produced by outgassing, the mass of the nucleus imposes a limit for the maximum mass of the ejected meteoroids (Jenniskens, 1998; Gundlach et al. 2015). In the lowest part of the mass distribution, small particles usually do not provide evidence for a cut-off in the distribution for the range of masses associated with visual meteors. Only in one case, radar observations find evidence for a low mass cut-off in the Geminid stream. The Ondrejov meteor radar has noticed the absence of echoes below magnitude +8 (Pecina, 2003). Such a radar magnitude corresponds to an estimated mass around $10^{-5}$ grams using equation (6). We will come back to this topic in the discussion section.

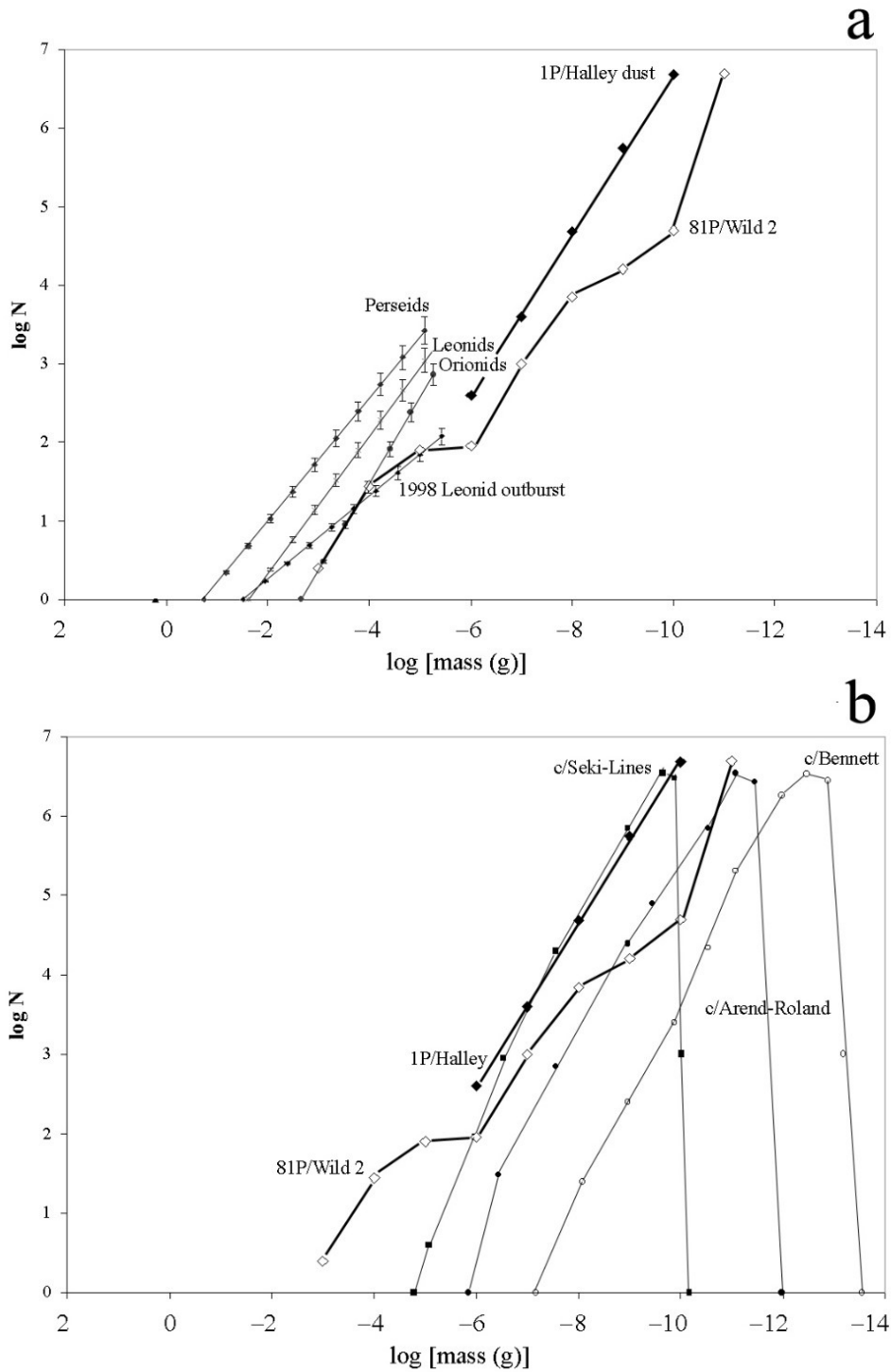

Figure 2. Number of ejected particles as a function of their mass. a) Spacecraft data for 1P/Halley and 81P/Wild 2 (Mc Donell et al, 1987; Green et al., 2004) compared to typical meteoroid mass distributions. The error bars are plotted assuming a 5% uncertainty (proportional to the number of particles of every mass present in the distribution). b) Same spacecraft data as in a, but now compared with remote determinations of the dust in the coma of comets Seki-Lines, Arend-Roland and Bennet (data from Delsemme, 1983).

Table 1. Incoming mass from meteoroid streams during annual encounters, outburst or storms. For "annual" we average the normal flux of particles intercepted by the Earth every year. That annual flux is nearly constant, but dust trails are sometimes intercepted by the Earth producing higher rates than the normal. When the activity overpasses the annual rate, these events are called "outbursts" or, if the meteor activity is higher than 1000 meteors/hour, "storms". It is defined as the nature of the meteor activity in column 2 for coherence with the literature. Note that the activity period or duration of the annual shower, outburst or storm is defined by the solar longitude interval (position of the Earth in its orbit). The spatial number densities ($\rho_{6.5}$) for particles producing meteors brighter than +6.5 magnitude are an estimation of the number of visual meteoroids included in a cube of 1000 km edge length. References: [1] IMO, 2000; [2] Jenniskens, 1995; [3] Rendtel et al., 1998; [4] Arlt et al. (1999), [5] Jenniskens, 1994; [6] Trigo-Rodríguez et al. (2013) [7] Arlt & Brown, 1999; [8] Trigo-Rodríguez et al., 2001; [9] Arlt et al., 1999b; [10] Trigo-Rodríguez et al., 2004a; [11] Rainer et al., 2002 [12] Dubietis, 2003, [13] Blaauw et al. (2015), and [14] Moreno-Ibáñez et al. (2017).

| Meteoroid Stream | Nature | $V_g$ (km/s) | R | Solar Longitude Interval (°) | Incoming Mass (kg/activity period) | Max. ZHR (meteors/hour) | $\rho_{6.5}$ (n/$10^9$ km$^3$) | REF. |
|---|---|---|---|---|---|---|---|---|
| Bootids | Annual | 18 | ~2.5 | 95.0-96.4 | 6 ± 2 | Variable (5 ± 3) | 12 ± 5 | [1] |
| | 1916 outburst | | 1.7 | 95.6 | 100 ± 30 | 300 ± 80 | 800 ± 200 | [2-4] |
| | 1998 outburst | | 2.2 | 95.7 | 145 ± 35 | 300 ± 50 | 800 ± 130 | [3-4] |
| Delta Aquarids | Annual stream | 43 | 2.5 | 115-132 | 3 ± 1 | 12 ± 3 | 10 ± 2 | [1,5] |
| Eta Aquarids | Annual stream | 66 | 2.4 | 30-56 | 7 ± 2 | 25 ± 10 | 40 ± 10 | [1,5] |
| Geminids | Annual stream | 36 | 2.2 | 250-266 | 75 ± 25 | ≈150 | 50 ± 15 | [5] |
| Giacobinids | 1933 storm | 20 | 3.2 | 196.2-196.4 | 550 ± 150 | 10000 ± 2000 | 84000 ± 22000 | [2] |
| | 1946 storm | | 3.2 | 196.2-196.4 | 700 ± 200 | 12000 ± 3000 | 84000 ± 22000 | [2] |
| | 1985 outburst | | 3.4 | 194.3-194.8 | 200 ± 50 | 700 ± 100 | 4500 ± 1500 | [2] |
| | 2011 outburst | | 3.0 | 194.9-195.2 | 950 ± 150 | 419 ± 59 | 2700 ± 700 | [6] |
| Leonids | Annual | 71 | 2.4 | 220.0-247.0 | 2 ± 0.5 | 30 ± 10 | 15 ± 3 | [1] |
| | 1966 storm | | 2.9 | 234.3-234.6 | 8.5 ± 2.5 | 15000 ± 3000 | ≈10000 | [2,7] |
| | 1998 outburst | | 1.2/2.0 | 234.0-235.5 | 1700 ± 600 | 300 ± 100 | 200 ± 40 | [7,8] |
| | 1999 storm | | 2.2 | 235.2-235.4 | 25 ± 8 | 3700 ± 100 | 5400 ± 1200 | [8,9] |
| | 2002 storm | | 2.2 | 236.5-236.7 | 30 ± 10 | 4500 ± 100 | 6600 ± 1900 | [10,11] |
| Orionids | Annual | 66 | 2.5 | 200-216 | 1.6 ± 0.5 | 25 ± 5 | 20 ± 4 | [1,5,12] |
| Perseids | Annual shower | 59 | 2.2 | 105-155 | 150 ± 45 | 120 | 100 ± 15 | [1,5,13] |
| | 1991 outburst | | 1.9 | 138.8-139.0 | 250 ± 50 | 500 ± 100 | 500 ± 150 | [2,5] |
| Quadrantids | Annual | 43 | 2.1 | 282.5-283.9 | 14 ± 4 | Variable (95 ± 15) | 150 ± 30 | [1,5] |
| Taurids | Annual | 28 | 2.3 | 190-250 | 155 ± 50 | 7 ± 3 | 15 ± 5 | [1,2] |
| Ursids | Annual | 35 | 2.2 | 265-277 | 11 ± 3 | 15 ± 5 | 25 ± 8 | [1,2] |
| | 1986 outburst | | 2.2 | 270.0-270.6 | 35 ± 10 | 160 ± 40 | 220 ± 45 | [2,5] |
| | 2014 outburst | | 1.8 | 271.6-272.1 | 15 ± 5 | 45 ± 19 | 70 ± 20 | [14] |

## 3.2. IS OVERABUNDANT INTERSTITIAL SODIUM PLAYING A ROLE IN JOINING COMETARY AGGREGATES?

From the analysis of ablation columns of cometary meteoroids using meteor spectroscopy, it has been reported that Na is mainly contributing to the meteor light during the first stages of ablation (Borovička et al., 1999; Trigo-Rodríguez et al., 2003, 2004). Additionally, it has been suggested that the sodium abundance can be higher than the solar abundance (Trigo-Rodríguez et al., 2004), which was later confirmed by the *Rosetta* study of comet 67P/Churyumov-Gerasimenko (Altwegg et al., 2015). This finding can be explained, because Na could be aggregated into another reservoir (organics and/or ices) in the outer parts of the nebula where comets formed. Our recent discovery of an ancient cometary clast in the interior of a CR chondrite reinforces such a scenario (Nittler et al., 2019). In fact, it is known that during the early stages of formation of the Solar System intense solar UV radiation from the evolving Sun depleted the content of some lithophile elements, such as Na, K or Mn in chondritic meteorites (Wasson & Kallemeyn, 1988; Wasson, 1988; Despois, 1992). After that, rocky bodies that formed in the inner Solar System suffered differentiation and degradation processes that probably also eroded volatile elements (Ehrenfreund et al., 1997). In fact, today the process of sodium depletion of the inner Solar System still continues, due to the solar wind that removes this element of young comets and tenuous atmospheres where it is accumulating by vaporisation of meteoroids. For these reasons, sodium abundances are expected to be higher in unprocessed bodies (i.e. comets) formed in the outer region of the protoplanetary disc (Despois, 1992). We propose a simplistic model in which Na became incorporated into the volatile-rich interstitial fine-grained matrix that is consistent with recent *Rosetta* observations (Schulz et al., 2015). Such a model is also consistent with our meteoroid bulk chemical data for the main rock-forming elements in which Na was proved to be overabundant (Trigo-Rodríguez et al., 2004; see also our Fig. 3)

Sodium has been also widely observed in cometary comae, but there is an intrinsic difficulty of establishing the origin of Na and its relative abundance from that data. Cremonese et al. (2002) measured a Na/Si atomic ratio of $\sim 3\times10^{-5}$ in the coma of comet Hale-Bopp that was explained as being produced from sputtering of the particle surfaces only. The sodium present in the tail of some comets forms a well-defined Na-tail (Cremonese et al., 1997; Cremonese and Fulle, 1999). These authors suggested that this sodium could be produced in a near-nucleus region, probably by cometary degassing or through an extended source, e.g. by the break-up of Na-bearing molecules, ions or dust particles. Detailed studies are required in order to infer the exact mechanism of neutral sodium release from dust particles. If sodium was embedded into organic material or ices associated with low temperature evaporating phases, as suggested by meteor spectra results, the amount of sodium released from some comets could be easily explained.

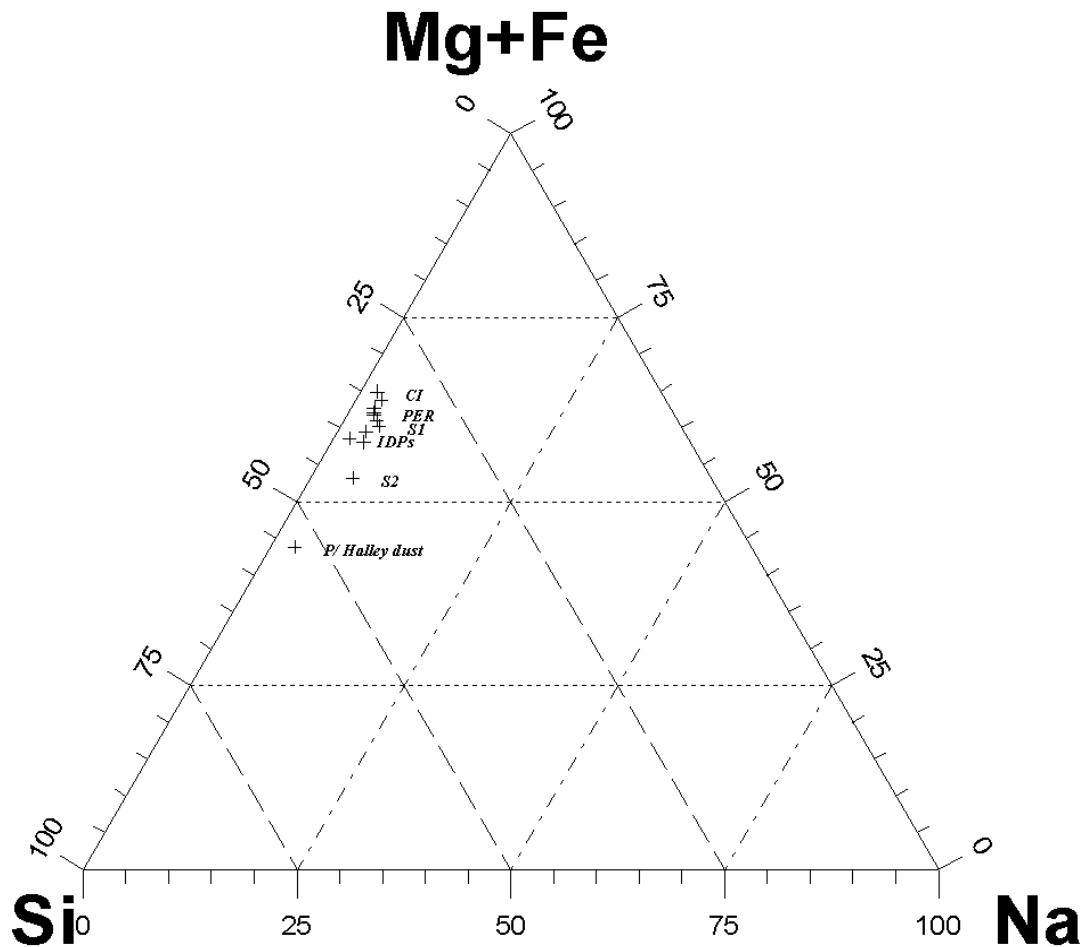

Figure 3. In this ternary diagram are compared the Mg+Fe, Si and Na abundances in the selected sample of cometary particles with this found for the CI chondrites, aggregate IDPs and 1P/Halley dust abundances (data from Trigo-Rodríguez et al., 2003). Additional labels are: a sporadic meteoroid (SPO2) and an averaged value for five Perseids (PER).

4. DISCUSSION.

4.1. MASS DISTRIBUTION AND DELIVERY OF COMETARY METEOROIDS.

The mass integration of the different streams allows us to compare the magnitude of the delivery of cometary matter to the Earth for every meteoroid stream (Table 1). Meteors produced by high encounter velocity streams are more spectacular, because also tiny meteoroids produce observable meteors. Despite this, those produced by low-velocity streams can deliver cometary material very efficiently. According to our results, the annual meteoroid streams deliver tens of kilograms during their periods of activity. In any case, meteor outbursts or storms produced when the Earth crosses fresh dust trails deliver several orders of magnitude more mass than the annual streams. Not all meteor outbursts during the last decades were produced by high-velocity material. For example, on June 27-28, 1998, an unexpected outburst produced by meteoroids ejected from comet 7 P/Pons-Winnecke delivered ~150 kg of cometary matter in an effective time interval that lasted an estimated 7 hours (Arlt et al., 1999; Asher & Emel'yanenko, 2002). Another recent case occurred on October 8, 1998, when an outburst was produced by Giacobinni-Zinner meteoroids (Simek & Pecina, 1999). The incoming meteoroid velocities for these streams are 18 and 20 km/s, respectively, and they are probably delivering important amounts of IDPs. Nowadays, it is possible to forecast reasonably well the appearance of perturbed dust trails that only in rare occasions intercept the Earth's orbit. The study of low geocentric velocity minor showers capable of producing unexpected outbursts is especially relevant. Between these cometary streams, the June-Bootids (Pons-Winneckids) and the October Draconids (Giacobinids) deserve special attention. These cometary streams are subject to important planetary perturbations (Jenniskens, 1995; 1998; Asher & Emel'yanenko, 2002) that introduce a high degree of difficulty to predict their returns, but that are feasible for modern computational techniques. Future campaigns to collect IDPs in the lower stratosphere can be coordinated by considering the time expected for these outbursts and the time required for these ablation products to settle through the stratosphere.

Visual, photographic and video techniques provide information on the mass distribution of cometary streams in the mass range from several kilograms to $\sim 10^{-6}$ g. We have compiled the range of masses and sizes derived from visual and photographic meteor data for cometary meteoroids in Table 2. However, what is the minimum size of the particles reaching the Earth from cometary meteoroid streams? In order to answer this question, Pecina (2003) provided additional information on very small particles registered with the Ondrejov Observatory meteor radar (Czech Republic). The radar limiting magnitude of +9 corresponds to masses ranging from $10^{-7}$ to $10^{-8}$ g, depending on the velocity (Pecina, 2003). No low mass cutoff is detectable with radar for the known cometary streams except for the Geminid stream. This indicates that extremely small particles with masses below $10^{-7}$ g to $10^{-8}$ g (and typical cross-sections of survival) are continuously reaching the upper atmosphere from cometary streams. Radar counts of meteors fluxes are performed thanks to the scattering of radio waves from meteor trails that might be achieved through radar observations by studying the waves reflected or scattered from the ionized meter trails (see e.g. Pecina, 2015).

The low-mass cut-off inferred for the Geminid stream is especially interesting, because the parent body of this stream is 3200 Phaeton, an intriguing object discovered by the Infrared Astronomical Satellite (IRAS) in 1983. Fred L. Whipple was the first to point out the similarity between Phaeton's orbit and the Geminid stream (Whipple, 1983). Whether the Geminid stream was produced by cometary activity of asteroid 3200 Phaeton is currently a matter of debate. However, meteor spectroscopy has provided recent clues. Licandro et al. (2010) proposed that this asteroid originated from the Pallas asteroid family, being probably a piece of disrupted asteroid 2 Pallas. On the other hand, on the basis of reflectance spectroscopy DeLeón et al. (2010) found that 3200 Phaeton exhibits a reflectance spectrum similar to its presumably parent asteroid 2 Pallas. Consequently, the parent body of the Geminid meteoroid stream could be more a transitional asteroid that experienced cometary-like activity in the past. It seems obvious that the densely populated Geminid stream cannot be easily explained by a punctual asteroidal collision. For example, the Na abundance of these meteoroids suggest different ejection ages like those expected for cometary activity (Borovička et al., 2005). Some Geminid spectra are poor in Na as well (Trigo-Rodríguez et al., 2004d), being far from the expected Na-rich aggregates expected in cometary materials (Trigo-Rodríguez and Llorca, 2007).

Table 2. Typical range of masses and sizes derived from visual and photographic meteor data for cometary meteoroids compared with data derived from the study of cometary comae and laboratory analyses of IDPs. In some cases, an asterisk indicates that the minimum mass has been extended to the limiting meteor magnitude detected by radar observations. The size is calculated assuming a spherical shape for the meteoroid for different assumed densities ($\delta$, in g/cm$^3$). Sources: [1] Rietmeijer (2002), [2] Fulle (1988), [3] Sarmecanic et al. (1997), [4] Sarmecanic et al. (1998), [5] Fomenkova et al. (1999), [6] Spanish Meteor Network data, [7] Visual meteor database of the International Meteor Organization (IMO, 2000), [8] Hajduk (1987) [9] Murray et al., (1999), and [10] Pecina and Pecinová, D. (2004)

| | Mass (g) | Size (cm) [$\delta$=0.1] | Size (cm) [$\delta$=1] | Size (cm) [$\delta$=3] | Sources |
|---|---|---|---|---|---|
| **Coma measurements** | | | | | |
| P/Halley's coma | $10^{-5}$ to $10^{-20}$ | - | $3 \cdot 10^{-5}$ to $3 \cdot 10^{-10}$ | $2 \cdot 10^{-5}$ to $2 \cdot 10^{-10}$ | [1] |
| Arend-Roland 1957III | $10^{-6}$ to $10^{-9}$ | - | $10^{-2}$ to $10^{-3}$ | - | [2] |
| C/1996B2 Hyakutake | $10^{-17}$ to $10^{-20}$ | - | $2 \cdot 10^{-6}$ to $10^{-7}$ | - | [3] |
| 109P/Swift-Tuttle | $10^{-17}$ to $10^{-22}$ | - | $10^{-6}$ to $10^{-8}$ | - | [4] |
| 55P/Tempel-Tuttle | $10^{-18}$ to $10^{-20}$ | - | $10^{-6}$ to $10^{-7}$ | - | [5] |
| **Laboratory data of IDPs** | | | | | |
| Aggregate IDPs | $10^{-9}$ | - | $1.2 \cdot 10^{-6}$ | - | [1] |
| Cluster IDPs | $5 \cdot 10^{-7}$ | - | $10^{-5}$ | - | [1] |
| **Meteoroid streams (range of meteoroid masses producing visual meteors)** | | | | | |
| Quadrantids | 1 to $10^{-5}$ | 1.6 to 0.035 | 0.8 to 0.016 | 0.5 to 0.011 | [6, 7] |
| Lyrids | 1 to $10^{-5}$ | 1.6 to 0.035 | 0.8 to 0.016 | 0.5 to 0.011 | [6, 7] |
| π Puppids | 10 to $10^{-3}$ | 3.5 to 0.015 | 1.6 to 0.8 | 1.1 to 0.05 | [6, 7] |
| η Aquarids/Orionids | 0.1 to $10^{-6}$ | 0.8 to 0.015 | 0.4 to 0.008 | 0.2 to 0.005 | [6, 7, 8] |
| June Bootids | $10^2$ to $10^{-3}$ | 8 to 0.16 | 4 to 0.08 | 2 to 0.04 | [6, 7] |
| α Capricornids | $10^2$ to $10^{-4}$ | 8 to 0.08 | 4 to 0.04 | 2 to 0.02 | [6, 7] |
| Giacobinids* | $10^2$ to $10^{-5}$ | 8 to 0.04 | 4 to 0.02 | 2 to 0.01 | [6, 7] |
| Taurids | $10^4$ to $10^{-5}$ | 35 to 0.04 | 16 to 0.02 | 11 to 0.01 | [6, 7] |
| Leonids* | $10^4$ to $10^{-9}$ | 35 to 0.002 | 16 to 0.001 | 11 to 0.0005 | [1, 6, 7, 9, 10] |
| Geminids* | 10 to $10^{-6}$ | 4 to 0.02 | 2 to 0.01 | 1 to 0.005 | [6, 7, 10] |

Considering the Geminid geocentric velocity, the observed low mass cut-off corresponds to particles with a mass ~$10^{-5}$ g. We can compare this minimum mass with those associated with other cometary meteoroids in the visual range. For example, a +6 magnitude Leonid is produced by a meteoroid of ~$4 \cdot 10^{-6}$ grams and a Perseid of identical magnitude comes from an ~8·10⁻⁶ gram particle. Taking into account radar observations, the difference is still larger, because these cometary showers have particles with a mass of $10^{-7}$ g producing +8 magnitude meteoroids unseen at naked eye. Our estimated range of diameters for cometary particles assuming spherical shapes and typical bulk densities are given in Table 3.

Table 3. Expected diameter (in millimeters) for cometary particles assuming spherical shapes and typical densities. For comparison the second column shows the visual magnitude of the meteor (obtained from equation 6) calculated for two different geocentric velocities. Note the difference in visual magnitude as consequence of the different geocentric velocity.

| Density (g/cm³)→ Mass (g) ↓ | Visual Magnitude | | 3.5 | 2.0 | 0.7 | 0.1 |
|---|---|---|---|---|---|---|
| | 15km/s | 60km/s | | | | |
| 0.001 | +7 | +1 | 0.09 | 0.11 | 0.16 | 0.30 |
| 0.004 | +6 | 0 | 0.13 | 0.16 | 0.22 | 0.42 |
| 0.011 | +4.5 | -1.5 | 0.19 | 0.22 | 0.31 | 0.59 |
| 0.082 | +2.5 | -3.5 | 0.37 | 0.43 | 0.61 | 12.0 |
| 0.224 | +1.5 | -4.5 | 0.51 | 0.6 | 0.85 | 16.0 |
| Mean characteristics | | | Non-porous chondritic (silicates+carbon) | Dense IDPs | Clustered porous IDPs | Fluffy and fine-grained IDPs |

The most likely explanation for the cutoff in the Geminid stream is the disappearance of particles below this mass by some physical process. By looking at the stream's age and peculiar orbit, we can find out a likely cause. This stream has a short orbital period of 1.5 years, an inclination of only 23º and a perihelion distance of only 0.14 A.U. Since this stream was formed in a short period orbit more than a thousand years ago (Williams & Wu, 1993), their meteoroids have been subjected to collisions with interplanetary dust and heating during their close approaches to the Sun (Beech, 2002). In addition, the Geminid stream is subjected to Poynting-Robertson (P-R) drag, which removes the smallest particles from the stream over shorter timescales than the larger ones. Such a process substantially depletes the stream, preferentially for small particles (see e.g. Messenger, 2002). The P-R drag removal is a significant effect to consider because seems to be dominant over collisional processes. If collisions were so efficient at destroying cometary meteoroids on the thousand-year timescale, we would find few particles with long term exposure to the solar wind, which has a very shallow penetration depth, among the interplanetary dust collected from the stratosphere. The P-R drag should remove efficiently meteoroids from their streams as the small IDPs collected by NASA from the Earth's stratosphere exhibit implanted solar wind noble gases indicating that particles with exposure ages of tens of thousands of years are common (Kehm et al., 2006). Additional evidence can be found from larger micrometeorites collected from polar ices that also support long exposures to solar wind (Olinger et al., 2000).

Consequently, over long time scales the P-R drag and the disruption of the smallest meteoroids explains the absence of small meteoroids in evolved streams like the Geminids. It is well known that collisional processes are crucial in the evolution of cometary dust in the inner Solar System (see e.g. Steel & Elford, 1986; Hughes, 1993; Mukai et al. 2001). In fact, the so-called β-meteoroids are dust particles with sizes below 1μm leaving the Solar System in hyperbolic orbits produced by collisions (Mann et al., 2004). In the previously mentioned article was pointed out that the number of particles with masses m<$10^{-6}$ g is greatly reduced relative to the largest ones as observed for the Geminid stream. Trigo-Rodríguez et al. (2005) reported the detection of unusual orbits of Leonid meteoroids probably produced by collisions with interplanetary dust that reduced the original meteoroid sizes and heliocentric velocity. Such a physical process, also including mutual collisions between meteoroid stream members (Williams et al., 1993), also contributes to decrease the number density of dust trails.

From all these techniques to infer the mass distribution of meteoroids reaching the top of the atmosphere we can conclude that a relative small fraction of the IDPs should have their origin in comets and should be collected just after Earth's crossing of young dust trails. When meteoroid geocentric velocities are higher than about 20 km/s the encounter produces intense meteor activity in the optical range, but lower geocentric velocity encounters can be unnoticeable by optical detectors, causing the main delivery of IDPs. As fluffy particles associated with dense and young dust trails have significant differences in relative velocity, it might cause an efficient fragmentation by mutual collisions when relative speeds exceed ~1 m/s (Güttler et al. 2010; Bukhari Syed et al. 2017). Consequently, we predict that the collection of IDPs during such encounters with the Earth might be especially fruitful.

If we consider the amount of material delivered by the main meteoroid streams given in Table 1, we could reach a value of about 0.425 tons/year. Obviously our Table only includes 11 out of 112 well-established meteor showers according to the IAU Meteor Data Center (MDC, 2021). It means that about one hundred additional cometary streams are active (mostly producing minor showers), and the Earth receives also a significant flux of sporadic meteoroids from comets not computed here. In any case, assuming that the computed value is a lower limit for the densest streams, the total annual contribution could be about two or three orders of magnitude over that value (let's say ~500 tons/year). Certainly, the current annual contribution is higher in years where there is a storm or an outburst, but on average probably not more than a factor of 3 or 4 greater since there isn't even one storm or outburst every year. Even when our ultimate goal was not inferring the global flux considering all sources, our results are below the assessment of the current dust accretion rate onto the Earth. For example, a more comprehensive study by Drolshagen et al. (2017) found that flux to be 32 tons/day (or $11.7\times10^3$ tons/year). We also remark that the bulk of this contribution is formed by particles in the mass range from $10^{-6}$ to $10^{-4}$ g (Figure 4 of Love and Brownlee, 1993), who computed the flux in this peak as being ~30,000 tons/year. Sekanina (1976) detailed study of meteors found that about a 16% of a total of 19,698 radio meteors were associated with cometary streams. Many of these meteoroid streams have not a clear parent comet because over relatively short timescales they tend to disintegrate (Sekanina, 1982, 1984).

Other authors suggest a highest flux contribution from comets. For example, Carrillo-Sanchez et al. (2016) suggested that Jupiter Family Comets (JFCs) contribute (80 ± 17)% of the total input mass (43 ± 14 tones/day or $15.7 \times 10^3$ tons/year). Our results are not favoring such high values, pointing that perhaps the annual contribution of cometary stream particles to the Earth is smaller than previously thought, probably a negligible fraction of the total mass of interplanetary dust accreting onto the Earth every year. If our approach is correct there is a clear conclusion, that comets make a minor contribution to the IDPs accreting onto the Earth. In any case, we are always talking about the flux inferred from visual meteors, and there is a clear bias for meteoroid streams producing slow-moving meteors that might be unnoticed and contributing more significantly to the total flux. In addition, the mass flux delivered by a meteor storm could be higher due to the presence of extremely fine-grained organic-rich meteors that are not producing bright meteors in the visible range (see Trigo-Rodríguez et al., 2013) for further evidence of that possibility in the last October Draconids storm. Consequently, we think that our results are not necessarily in controversy with other authors, but seem to require other pathways in which cometary materials reach the Earth. In fact, we discovered that the stream meteoroids are eroded over short timescales by collisions with interplanetary dust (Trigo-Rodríguez et al., 2005), or being perturbed in planetary close approaches (Trigo-Rodríguez et al., 2006) so there are ways to increase the amount of IDPs arriving from comets in "sporadic" orbits. In fact, we also think that our lower flux computed also emphasizes the need to count with all sources of cometary materials, particularly a detailed model to compute the different range of sizes of the meteoroids reaching the Earth from sporadic origin (see e.g. Staubach et al., 1997; Landgraf et al., 2000) if we really want to compute the total flux of materials arrived to Earth from comets.

## 4.2. PREATMOSPHERIC STRUCTURE OF METEOROIDS AND IDPs

The association between meteoroids and IDPs with comets is based on different criteria. Cometary meteoroids can be imaged from several stations and their orbits can be accurately linked with their parent comets. Although present campaigns to collect IDPs can be coordinated with coinciding collection with dust trails encounters, this cannot provide a 100% guarantee that the recovered samples are stemming from only one source. Messenger (2002) modeled the orbital evolution of four dust trails associated with comets in a geometry of low atmospheric entry speed to allow prediction of the collection times of particles from them. He noticed that a meteor outburst from comet 26P/Grigg-Skjellerup should produce the highest concentration of cometary particles at Earth's orbit during April 23-24 of 2003. That prediction moved NASA to conduct a "Timed Collection" that allowed the recovery of particles in the stratosphere directly arrived from the dust trail from comet 26P. Such recovered particles exhibited, unlike typical IDP, very short space exposures. In other words, they didn't experienced implantation of solar wind gases, neither evidence for solar flare tracks, being consistent with a recent release from 26P comet. Later on, another collection of "fresh" IDPs was conducted for 73P/Schwassmann-Wachmann 3 comet, also inferred from Messenger (2002) calculations. To achieve such a success, modelling the orbital evolution of very small particles considering P-R drag is clearly needed. The reason is that the small particles quickly separate from the main meteoroid stream (see e.g. Klačka and Williams, 2002), so they may intersect Earth when the main stream from an outburst does not, or vice versa.

Consequently, to develop criteria in order to know whether sampled IDPs are of cometary origin is extremely important. Love & Brownlee (1994) obtained a correlation between the peak heating temperature and the entry velocity that was later used by Joswiak et al. (2000) in order to infer the range of velocities associated with the surviving IDPs. Assuming that the velocity can be inferred from a stepped-He release curve and from the Love & Brownlee model, they found that meteoroids survive ablation in the velocity range between 11 to 29 km/s. Several cometary streams exhibit similar entry velocities. In the laboratories, the samples that are considered more representative of cometary dust are aggregate IDPs. This conclusion is based on their estimated high entrance velocities, high Mg-contents, presence of amorphous carbon, enrichment in deuterium, and isotopic anomalies in N and C in the carbon-phase (Brownlee et al., 1995; Rietmeijer, 1998; Wooden, 2002). However, by comparing the typical sizes of these cometary IDPs with the meteoroid sizes deduced in the present work, we have noticed a difference of several orders of magnitude. In the stratosphere, the largest IDPs of cometary origin are clusters of IDPs no larger than 100 μm in size or $\sim 5 \times 10^{-7}$ g in mass (Rietmeijer, 2002) but some cometary meteoroids are centimeter-sized objects. By studying the luminosity of flares in meteors, Simonenko (1968) found that the mean effective radius of the separated particles is about 70 μm.

There are two possibilities for explaining this. The first one is that the collected IDPs have precursors in the smallest particles present in meteoroid streams that are efficiently decelerated in the atmosphere (Love & Brownlee, 1994). The second possibility is that some collected IDPs are surviving fragments of larger meteoroids. We say this because the ablation process for slow meteoroids is not likely to be as efficient as that for fast meteoroids. This second alternative is not consistent with the noble gases found to be implanted in most small IDPs (Kehm et al., 2006). Consequently, the short heating associated with the peak temperature reached during the entry of very fluffy meteoroids would remove a low melting temperature interstitial material (organics?), but the entire meteoroid structure could collapse. As a consequence, smaller grains would be released, and the ablation will become more efficient. Fragmentation of cometary meteoroids in the upper atmosphere is also consistent with their low bulk densities (see e.g. Bronshten, 1983). According to this picture, highly porous aggregates with low melting temperature have interstitial phases that are "gluing" the silicate phases to form cometary meteoroids. The higher the porosity of the aggregate, the easier is its fragmentation in the atmosphere, because the cohesive (tensile) strength scales inversely with the porosity (San Sebastián et al 2020). During the disruption process, some particles acquire enough energy to be separated from the meteoric ablation vapors. The deceleration is very low for these particles in the upper atmosphere (Simonenko, 1968). The early release of Na and other volatile elements in meteors (Borovička et al., 1999) could be directly associated with the interstitial material (Trigo-Rodríguez et al., 2004c). If the melting or vaporization point of the interstitial phases is reached, this could lead to the consequent fragmentation of small particles, as was simplified in the dustball model (Hawkes & Jones, 1975). Consequently, while high velocity cometary meteoroids produce tiny ablation products, IDP-like fragments of low velocity cometary meteoroids would be surviving ablation.

### 4.3. OVERALL IMPLICATIONS FOR EARLY EARTH DELIVERY OF VOLATILES

Usually, the *Grand Tack Model* (Walsh et al. 2011) is used to explain the main features observed in the distribution pattern of planets, and the current distribution of minor bodies in the main belt, among other observational evidence (see e.g. Morbidelli et al., 2012). In such a scenario, we envision an early period in planetary evolution, in which the final setting of Jupiter and Saturn naturally increased the flux of fragile volatile-rich bodies crossing the terrestrial planets. Direct impacts and close approaches of these bodies with planets were far more usual at those remote ages, and delivered huge amounts of pristine materials to rocky planets.

In an early Earth scenario, being our planet subjected to a high-meteoroid-flux and these materials much shortly delivered from disrupted comets, we envision a significant chemical contribution. Even today, the significance of this continuous flux of extraterrestrial materials is out of doubt, having direct implication in atmospheric chemistry (Plane et al., 2018). Such cometary materials could have had significant influence in the Hadean and early Archean environment (Chyba and Sagan, 1992; Jenniskens et al., 2000). Nowadays these materials are mainly delivered as IDPs, but still a significant part (studied in this paper) comes from periodic meteoroid streams. In the ancient past, close approaches to the terrestrial planets could have produced other ways of delivery to our planet in shorter timescales with significant implications for the delivery of organics and volatiles to early Earth (Trigo-Rodríguez and Saladino, 2019; Martins et al., 2020). The mass distributions of particles released by outgassing from a comet are significantly different to those produced during catastrophic disruption by collisions, or tidally-induced disruptions (Jenniskens, 1995, 1998).

Given that large cometary aggregates are fragile (Blum et al., 2006; San Sebastián et al. 2020) and seem to involve organics and Na as part of the matrix sustaining the mineral grains (Flynn et al., 2013), it is obvious that meteoroid volatiles are efficiently eroded by solar irradiation in the long timescales in which meteoroids are finally reaching our planet. In fact, the survival of large meteoroids in meteor outburst and storms produced by young dust trails suggest that these particles could be more efficiently delivered during close encounters with terrestrial planets (Trigo-Rodríguez et al., 2013). To test this hypothesis, the geologic record can be used, but also the Moon could be an excellent place to search for clues about the consequences that the arrival of disrupted comets during the heavier bombardment period early in the Solar System could have had on our planet. The exploration of the Moon could find some of these impacts dated precisely and providing interesting clues to extrapolate the importance of such processes for our planet and their possible relevance for the appearance of life.

## 4.4. EXPECTED PEAK SIZES AND INNER STRUCTURES OF COMETARY DUST

The plethora of new observational data from the *Rosetta* mission allows us to estimate the most likely sizes of those dust particles that ultimately end up in the meteoroid stream. Blum et al. (2017) compiled multi-instrument observations by *Rosetta* and from the ground and showed that the peak mass of the ejected dust is most likely in the size range between 1 cm and 1 m. Blum et al. (2017) also showed that cm-sized particles are abundant on the surface of comet 67P/Churyumov-Gerasimenko. These particles, a.k.a. pebbles, can be associated with the building blocks of comets in the formation scenario of cometesimals in which a cloud of pebbles in the solar nebula was concentrated by the streaming instability (Youdin and Goodman 2005) and then collapsed due to a gravitational instability (Johansen et al. 2007). Gundlach et al. 2020 showed that a pebble makeup of the cometary nucleus can quantitatively predict the dust and gas emission rates of comet 67P around perihelion as well as the size distribution of the emitted dust from pebble sizes onward. The pebbles themselves are products of collisional accretion in the solar nebula and their size depends on a variety of parameters, e.g. the heliocentric distance (pebbles formed further away from the Sun are smaller than pebbles at smaller distances), the size of the constituent dust/ice grains (smaller grains form larger pebbles, due to their higher intrinsic stickiness), the dust-to-ice ratio (water ice is collisionally sticker than dust), or solar-nebula parameters (Lorek et al. 2018). The pebbles might be internally inhomogeneous, as proposed by Fulle et al. (2019; 2020a;b) to explain the cometary emission of small dust particles and the space between the pebbles might be filled with high-porosity primitive dust aggregates, which were accreted during the gravitational collapse, as was inferred by Fulle and Blum (2017) to explain Rosetta observations of high porosity dust. The materials forming primordial cometary aggregates can be also exemplified by the ancient clast discovered in the interior of a CR carbonaceous chondrite (Nittler el al., 2019).

Hence, young cometary trails should be dominated in cm-sized pebbles and clusters thereof up to the maximum size (~1 m) that is capable to escape the gravitational field of the nucleus (Gundlach et al. 2015). Older trails will be enriched in decay products of the clusters and pebbles and, thus, will contain more meteoroids smaller than 1 cm in size. In general, the estimated meteoroid sizes discussed in our paper are consistent with those modelled for the 67P/Churyumov-Gerasimenko dust trail, dominated by mm-sized particles (Soja et al., 2015). We expect a significant progress in our understanding of the fragmentation of meteoroids in space as direct measurements of the spatial density of dust trails by spacecraft is being achieved. In that sense, a recent progress is the study of cometary dust trails by the *Helios* spacecraft which used in situ dust sensors to infer the distribution of interplanetary dust of several cometary trails (Krüger et al., 2020).

## 5. CONCLUSIONS

Meteor science has developed extraordinarily in recent decades. New imaging techniques have been applied and are available to the study of meteor magnitude distributions that can provide direct reliable information on the mass distribution, and physical properties of cometary meteoroids (see e.g. Vida et al., 2021). Additionally, accurate multiple station observations of meteor storms and outbursts have been achieved and significant trajectory and orbital data can be deduced from them. These new techniques of studying cometary meteoroids can provide valuable clues to the nature and structure of cometary bodies and their fragments. All these new possibilities make the study of meteors complementary to laboratory analyses or in-situ missions, being able to provide information on some unknown processes occurring in the interplanetary medium or in the terrestrial atmosphere. Our approach developed here allows us to reach the following conclusions:

1) By comparing the mass distributions of cometary dust, we have shown that a gap between the largest collected cluster IDPs and cometary meteoroids exists. Typical sizes of cometary meteoroids extent into the cm regime (Table 2), a size for which meteoroids are not able to survive atmospheric entry. However, if fragmentation occurs at high altitudes, total ablation would be inhibited. In such a case, some of the collected IDPs would be fragments of low-velocity meteoroids.

2) Cometary disintegration products are compatible with the model of cometesimal formation (see Sect. 4.4). However, space-weathering processes are certainly altering the meteoroid mass distributions during long stays in the interplanetary medium. In-situ studies of comets with spacecraft and studies of meteor showers by using different techniques here exemplified can help us to constrain the mass distributions, compositions and physical properties of cometary materials reaching the Earth.

3) The by-product of space exposition of cometary meteoroids is a progressive loss of volatile elements. The aggregates are glued by a mixture of organics, and other elements that were probably transported by the stellar wind during the phases of cometary formation. These materials seem to play a key role in gluing the solid grains that form cometary meteoroids and explain the progressive disintegration of cometary meteoroids. As a consequence, meteoroid streams lose mass: decreasing the number density of large aggregates, but gaining smaller ones if the remaining fragments are not volatile, until their final disappearance in timescales of hundreds of thousands of years.

4) Concerning the delivery of moderately volatile elements and organics, our approach indicates that comets could have participated in massive delivery of sodium and organics to early Earth. This is particularly relevant in the context of the early period of high flux of transitional asteroids and comets through the inner solar system, as a significant number of close approaches to our planet could have created pathways of short-time scale delivery. Many aspects of the close encounters and arrival of these materials to our planet should be further investigated.

5) As a result of our measurements from the magnitude distributions of cometary meteoroid streams, it appears that the direct flux from meteoroid streams has a relatively minor contribution of cometary meteoroids accreting onto the Earth (~500 tones/year), so the sporadic source (dynamically evolved meteoroids from comets and asteroids) should be dominant.

**Data Availability Statement**

An additional supporting file with the data provided in this article will be deposited in a public repository or included as supplementary material to the article.

**Acknowledgments**

JMTR acknowledges financial support from the Spanish Ministry (PGC2018-097374-B-I00 research project grant. JB acknowledges the continuing support by the Deutsche Forschungsgemeinschaft and the Deutsches Zentrum für Luft- und Raumfahrt. Petr Pecina generously provided data on the meteoroid distribution in the range of masses belonging to radio meteors detected from the Ondrejov Observatory meteor radar (Czech Republic). We are grateful to George Flynn for his careful revision of our original manuscript, that really help us to improve our paper. Hans Betlem provided the updated version of the Dutch Meteor Society (DMS) orbital catalogue. Prof. Frans J. M. Rietmeijer provided valuable feedback to build Table 3.